\documentclass[a4paper]{jpconf}
\usepackage{graphicx}
\begin{document}
\title{C/O white dwarfs of very low mass: 0.33-0.5 M$_{\odot}$}

\author{P. G. Prada Moroni$^{1,2}$,  O. Straniero$^3$}

\address{$^1$ Physics Department ``E. Fermi'', University of Pisa,
largo B. Pontecorvo 3, I-56127, Pisa, Italy}
\address{$^2$ INFN, largo B. Pontecorvo 3, I-56127, Pisa, Italy}
\address{$^3$ INAF -- Osservatorio Astronomico di Collurania,
via Maggini, I-64100, Teramo, Italy}
\ead{prada@df.unipi.it}

\begin{abstract}
The standard lower limit for the mass of white dwarfs (WDs) with
a C/O core is roughly 0.5 M$_{\odot}$. In the present work we
investigated the possibility to form C/O WDs with mass as low as 0.33 M$_{\odot}$.
Both the pre-WD and the cooling evolution of such nonstandard
models will be described.
\end{abstract}

\section{Introduction}
It is by now a well-known and commonly accepted result of the stellar evolution
theory that stars of low and intermediate mass will end their lives as white dwarfs
(WDs). Since the pioneering papers by Chandrasekhar (1931) and Stoner (1932),
 it has been known that there must be an upper limit of
the mass of WDs. The current value for such a limit, now universally known as
 Chandrasekhar's mass, is  M$_{Ch}$= 1.456 (2/$\mu_e$)$^2$ M$_{\odot}$(where $\mu_e$ is
 the electronic mean molecular weight, Hansen, Kawaler \& Trimble 2004).
Below this critical mass there are essentially three different families of WDs 
concerning the chemical composition of their cores: the O/Ne/Mg, C/O and He. 
The upper mass limit for C/O WDs, which is still subject to some debate, since it is the 
aftermath of the evolution of a star with initial mass slightly lower than
M$_{up}$, the lowest mass for star that succeeds in igniting carbon burning 
before the asymptotic giant branch (AGB), it is around 1.05
M$_{\odot}$(Dominguez et al. 1999, Weideman 2000, Prada Moroni \& Straniero
2007, Catalan et al. 2008, Meng, Cheng \& Han 2008). 

On the other hand, the value for the minimum mass of WDs with a C/O core
is commonly believed to be around 0.5 M$_{\odot}$ (Weideman 2000, Meng et al. 2008). 
The theory of stellar evolution predicts that
an isolated single star with mass lower than about 0.5 M$_{\odot}$, the exact 
value depending on the chemical composition (0.46 and 0.50 for Z=0.04
and Z=0.0001, respectively, Dominguez et al. 1999, Girardi 1999), do not ignite the
 helium-burning and die as WDs with an helium rich core.
For these stars, during the red giant phase the electronic thermal
conduction and the emission of neutrinos, which cool the core, prevail on 
the heating due to the contraction caused by the accretion of fresh
helium processed by the hydrogen-burning shell. 
This is the main reason why WDs of mass less than about 0.5 M$_{\odot}$ 
are commonly believed to have a He-core.

On the other hand, as early as 1985 Iben \& Tutukov,
 in a very beautiful and instructive paper, showed that a star with
 an initial mass of 3 M$_{\odot}$ evolving in a close binary can produce
 a remnant of 0.4 M$_{\odot}$ with a sizeable C/O core.
A similar result was obtained 15 years later by Han, Tout \& Eggleton (2000),
who computed the evolution of close binary systems with several different 
parameters. They showed that in a binary system with 
the primary star of mass M$_1$= 2.51 M$_{\odot}$, the mass ratio
q=M$_1$/M$_2$=2 and the initial period P=2.559 d, the primary star succeeded to ignite 
the helium-burning. They did not followed the evolution of the remnant C/O WD, 
since their code met numerical problems when the primary star was as low as 0.33
 M$_{\odot}$ with a C/O core of 0.11 M$_{\odot}$.

In this paper we will investigate both a possible evolutionary scenario 
able to produce C/O WDs of low mass, that is M $< 0.5$ M$_{\odot}$, and their
cooling evolution.
We will show the results of detailed evolutionary 
computations performed with a full Henyey code able to follow consistently
the evolution of stars from the initial pre-main sequence to the final 
cooling phase of WDs. 
\section{The red giant phase transition}
The stellar models showed in the present work
 have been computed with an updated version of
 FRANEC (Prada Moroni \& Straniero 2002, Degl'Innocenti et al. 2008),
 a full Henyey evolutionary code.  
We adopted a metallicity Z=0.04 and an initial helium abundance Y=0.32 
suitable for the very metal-rich stars belonging to some open galactic
cluster, such as NGC6791.
\vspace{2pc}
\begin{figure}[h]
\begin{minipage}{16pc}
\includegraphics[width=16pc]{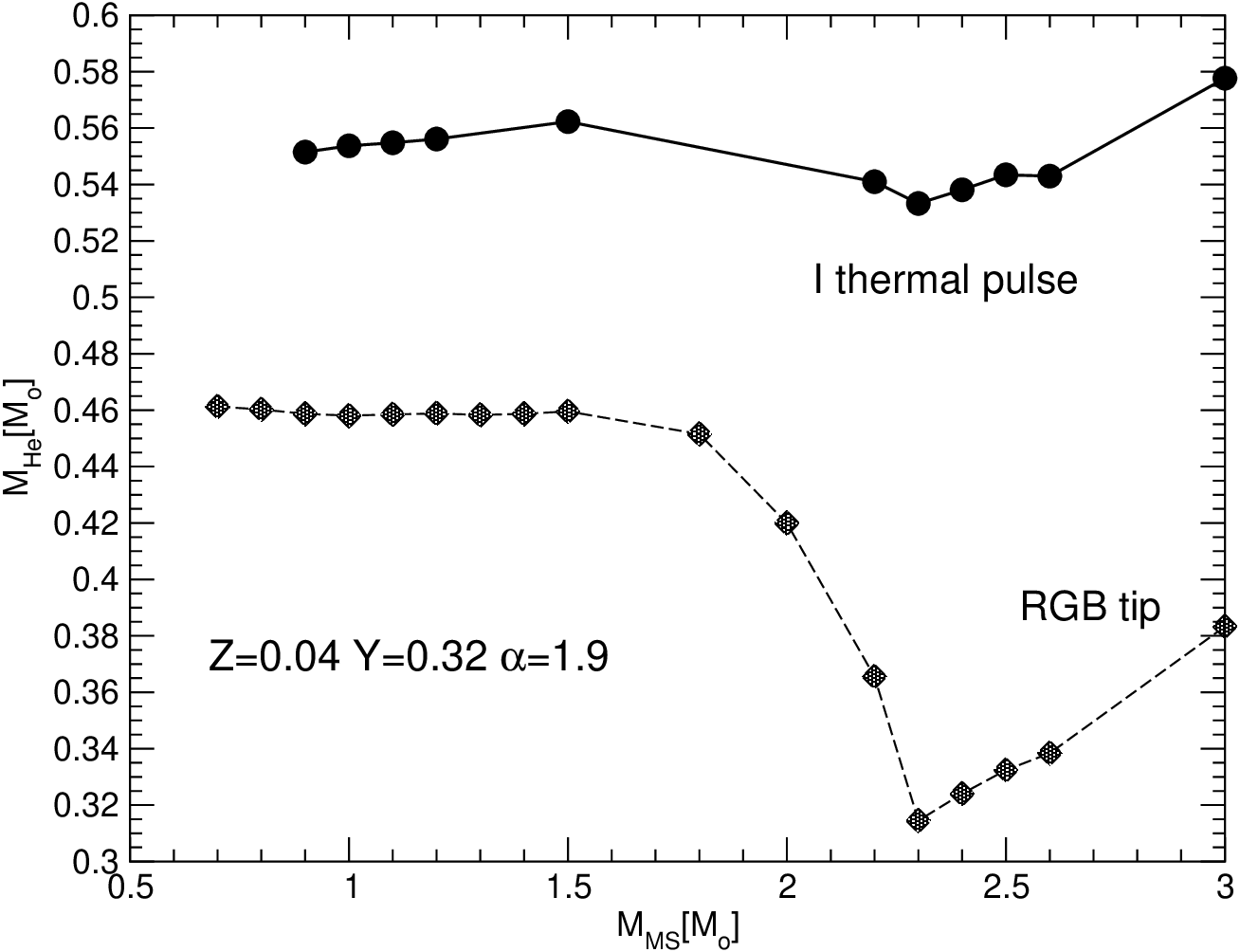}
\caption{\label{MheTip}Mass of the He-core at the tip of the RGB (diamond and
  dashed line) and mass of the hydrogen exhausted core at the first thermal
 pulse on the AGB (circles and solid line) as a function of the initial
 mass for stars with Z=0.04 Y=0.32.}
\end{minipage}\hspace{2pc}%
\begin{minipage}{16pc}
\includegraphics[width=16pc]{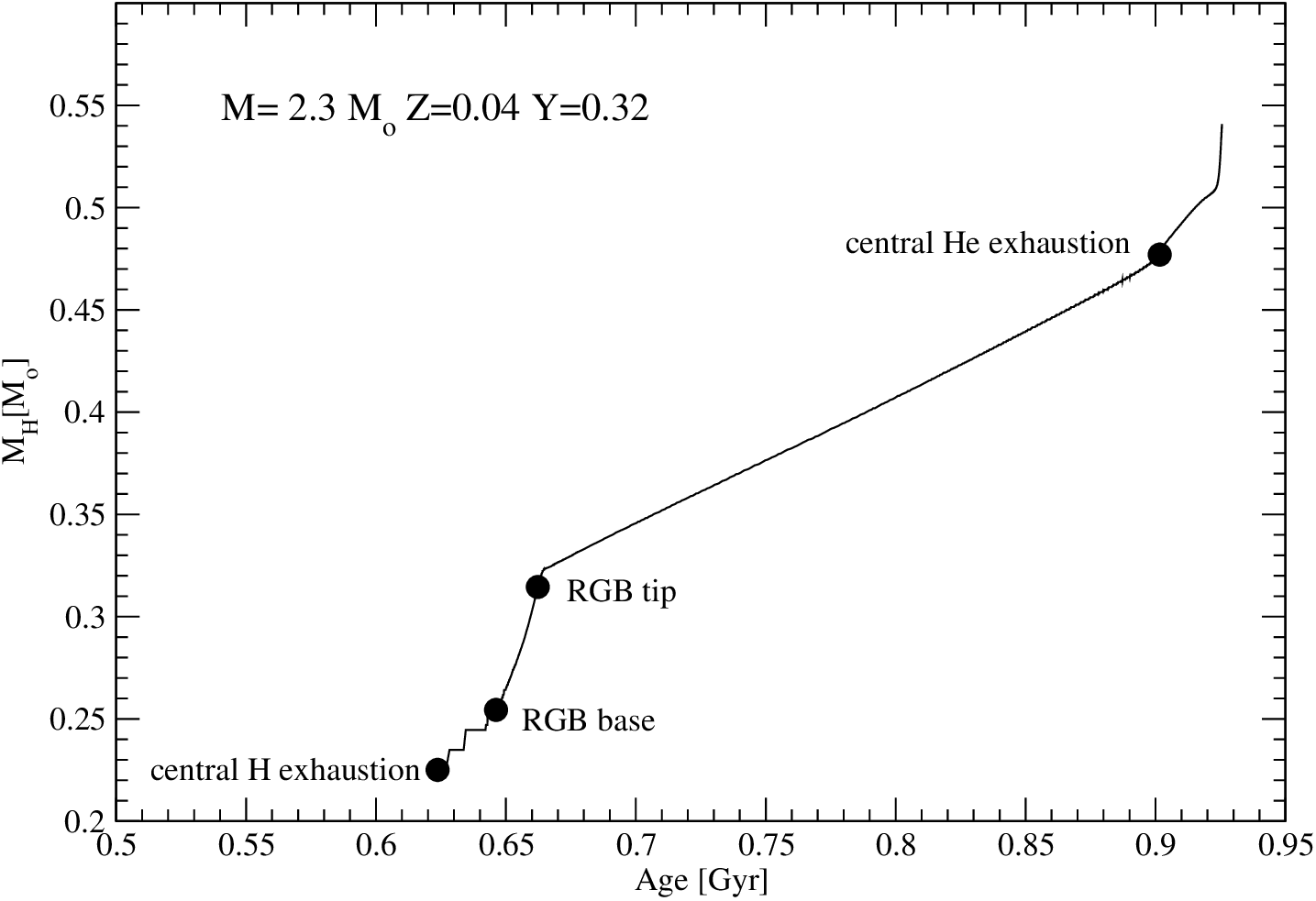}
\caption{\label{MheAge2p3} Mass of the hydrogen exhausted core as a function
  of the age for a star of M=2.3 M$_{\odot}$, Z=0.04 Y=0.32. The circles
mark: the central hydrogen exhaustion, the base and the tip of the RGB and 
the central helium exhaustion.}
\end{minipage}
\end{figure}

The figure \ref{MheTip} shows the mass of the hydrogen exhausted core M$_H$
 at the first thermal pulse on the AGB (upper curve) and the mass of the
 helium core M$_{He}$ at the tip of the red giant branch (RGB tip, lower curve)
 as a function of the initial mass. 
As it is well known, M$_H$ at the first thermal pulse is nearly constant
 around 0.55 M$_{\odot}$, the exact value depending on the chemical
 composition, for initial masses lower than 3 M$_{\odot}$. 
On the other hand, the behavior of the M$_{He}$ at the tip of RGB as a 
function of the initial mass is much less smooth than the previous one,
 in particular a deep and sharp minimum is present around 2.3 M$_{\odot}$ 
for this chemical composition (Castellani, Chieffi, Straniero 1992). 
Such a behavior is the consequence of the physical conditions 
present in the helium-core at the onset of the 3$\alpha$ nuclear reaction. 

For initial mass lower than 1.5-1.7 M$_{\odot}$, an electron-degenerate core 
develops at the beginning of the red giant phase, and the onset of the 3$\alpha$ 
occurs in a nuclear runaway, the so-called helium-flash (Hoyle \&
Schwarzschild 1955).
 For these stars, the M$_{He}$ at the tip of the RGB is almost constant, about
 0.46 M$_{\odot}$ for metal-rich stars and 0.50 M$_{\odot}$ for metal-poor
 ones, and is essentially produced during the hydrogen-shell burning phase.

From this point on, the larger the initial mass, the weaker the degeneracy of
 the core in the red giant phase, thus the weaker the He-flash and 
the lower the mass of the helium-core required for the 3$\alpha$ burning.
 When the initial mass is as high as about 2.3 M$_{\odot}$, 
the helium-burning ignites quiescently, without a
 flash. This explains the steep decrease of the M$_{He}$ at the tip of the RGB 
between 1.8 and 2.3 M$_{\odot}$.
For higher masses, the core in the red giant phase is not degenerate
 any more. From this point on, the larger the initial mass,
 the larger the M$_{He}$ at the tip of the RGB, as the larger the
 convective core during the previous main sequence phase.
Thus, the minimum mass of the helium-core M$_{He}$ at the onset of the
 3$\alpha$ occurs in the transition between these two regimes,
 around 2.3 M$_{\odot}$, reaching a value of 0.315 M$_{\odot}$. 

Notice that, as previously stated, this deep minimum is almost canceled at the 
first thermal pulse. In fact, the hydrogen burning shell continues to 
process hydrogen and move outward, then increasing M$_H$, during 
the central helium-burning phase. Moreover, it is a well established result
 of the theory of stellar evolution that the lower the mass of the He-core
 at the onset of the 3$\alpha$, the fainter the star and the longer
 the central helium-burning phase. Thus, as the initial mass of the
 helium-rich core decreases the hydrogen-burning shell will work for a longer
 time resulting in an larger $\Delta$M$_H$. This is the reason why star with
 initial mass lower than 3 M$_{\odot}$ enter the AGB phase with nearly the
 same M$_H$ (Dominguez et al. 1999). 

The first paper devoted to the study of the evolutionary characteristics of
stars belonging to this sharp and deep minimum in the 
mass of the He-core at the tip of the red giant phase was Iben (1967).
Since then, many studies have been focused on this transition, 
 named by Renzini \& Buzzoni (1983) red giant phase transition (hereafter RGB 
phase transition), 
 but a detailed discussion of such an impèortant feature is well beyond
 the aim of the present paper.
We will dwell only upon a feature of the evolution of a star with M= 2.3
M$_{\odot}$ important to our purpose.  
Figure \ref{MheAge2p3} shows the evolution of the mass of the
hydrogen-exhausted core M$_H$ for this transition mass. 
The circles mark the following evolutionary phases: the central-hydrogen exhaustion,
 the base and the tip of the RGB, respectively, and finally the central-helium exhaustion.
As one can easily see, when the star exhausts its central hydrogen, the mass 
of the He-core is already 0.225 M$_{\odot}$, which is more than the 70\% of
the value at the RGB tip. In fact, a sizeable convective core was developed
during the preavious main sequence phase. When the star reaches the base of
the RGB, the mass of the helium core is 0.254 M$_{\odot}$, which is about the
 80\% of the mass required to onset He-burning.
This means that, since a large fraction of the helium-core required
 for the ignition of the He-burning is already developed at the beginning
 of the red giant evolution, a star in the RGB phase transition, that is around 2.3
 M$_{\odot}$, that undergoes a very strong mass loss during the red giant phase can
 produce a C/O WD with a mass significative lower than 0.5 M$_{\odot}$.
\section{C/O white dwarfs of very low mass} 
\begin{figure}[h]
\begin{minipage}{16pc}
\includegraphics[width=16pc]{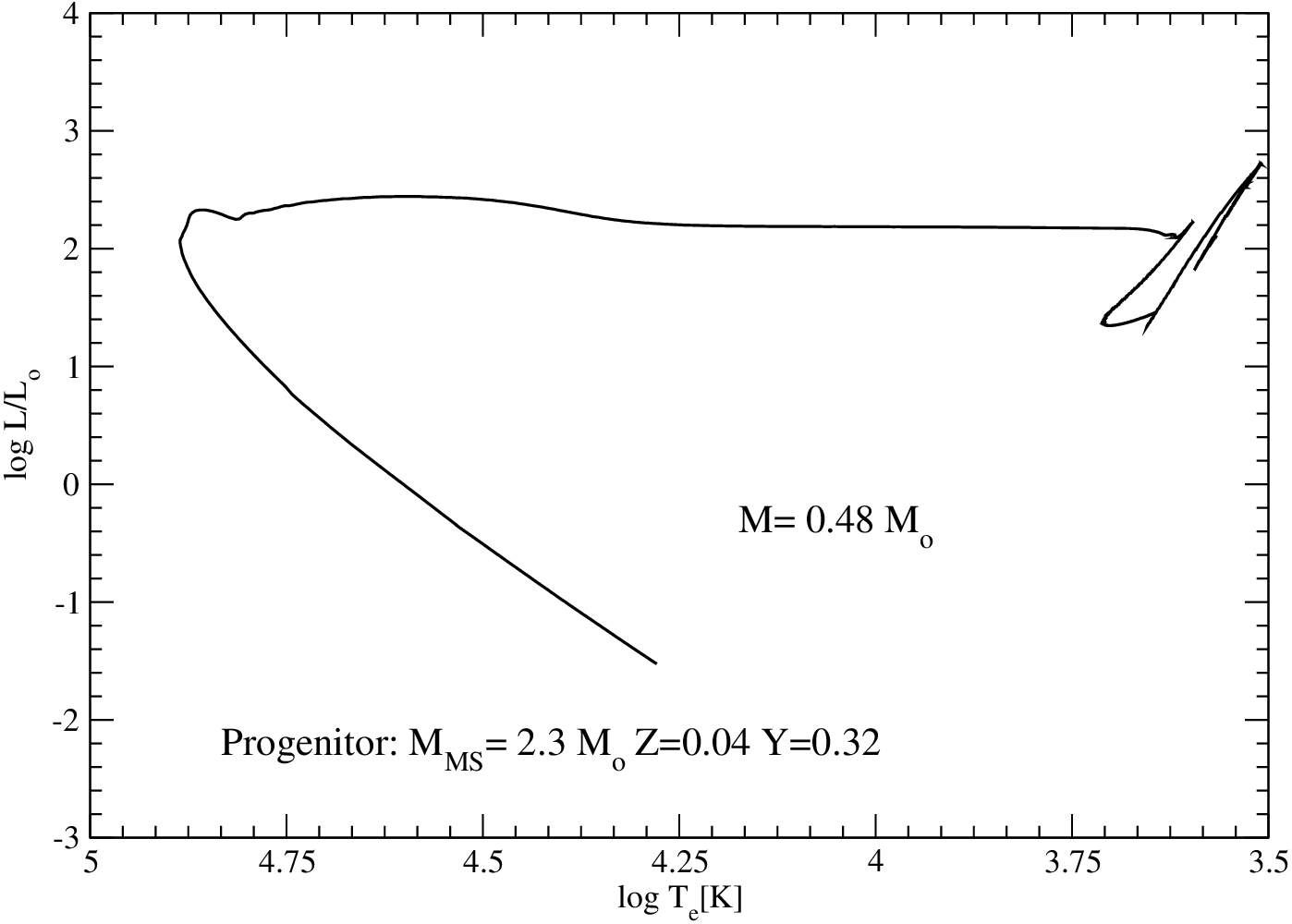}
\caption{\label{M0p48} Evolutionary track of a star of M=0.48 M$_{\odot}$ whose porgenitor of 
M= 2.3 M$_{\odot}$ underwent a strong mass loss episode during the red giant phase.}
\end{minipage}\hspace{2pc}%
\begin{minipage}{16pc}
\includegraphics[width=15pc]{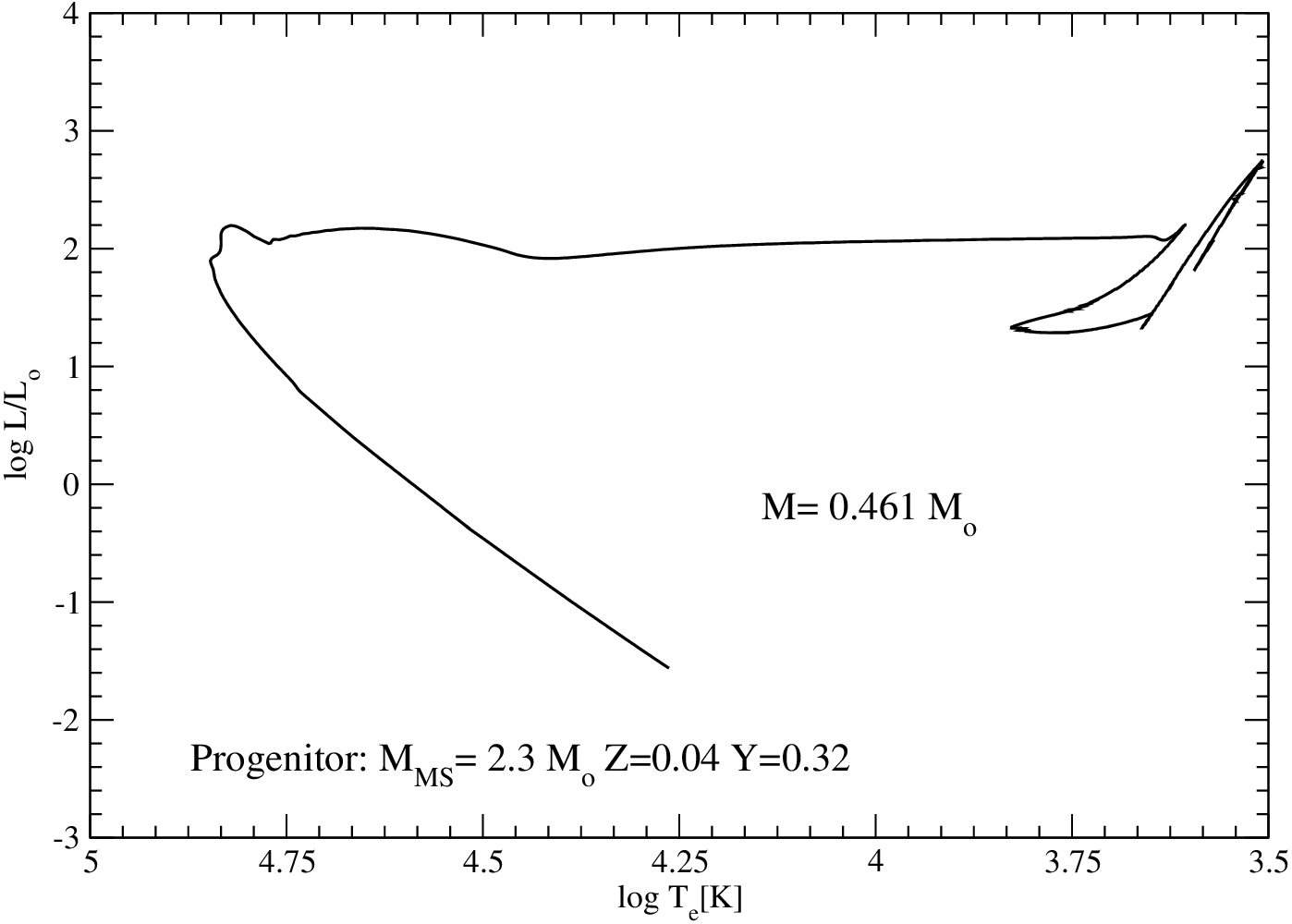}
\caption{\label{M0p461} Same as in figure \ref{M0p48} but for a remnant of mass
 M=0.461 M$_{\odot}$.}
\end{minipage}
\end{figure}%
In order to check such an idea and provide models of C/O WDs with mass lower than 0.5 M$_{\odot}$,
 we computed the evolution at constant mass of a star with M=2.3 M$_{\odot}$  
until the red giant phase. Then at about $logL/L_{\odot}$=1.34, when the
He-core was M$_{He}$= 0.2569 M$_{\odot}$, we turned on the mass loss, with a
constant rate of the order of 10$^{-7}$ M$_{\odot}$ yr$^{-1}$.
We stopped the mass loss once the desired final mass was obtained and then we 
followed the next evolution until the final cooling phase.

\begin{figure}[h]
\begin{minipage}{16pc}
\includegraphics[width=16pc]{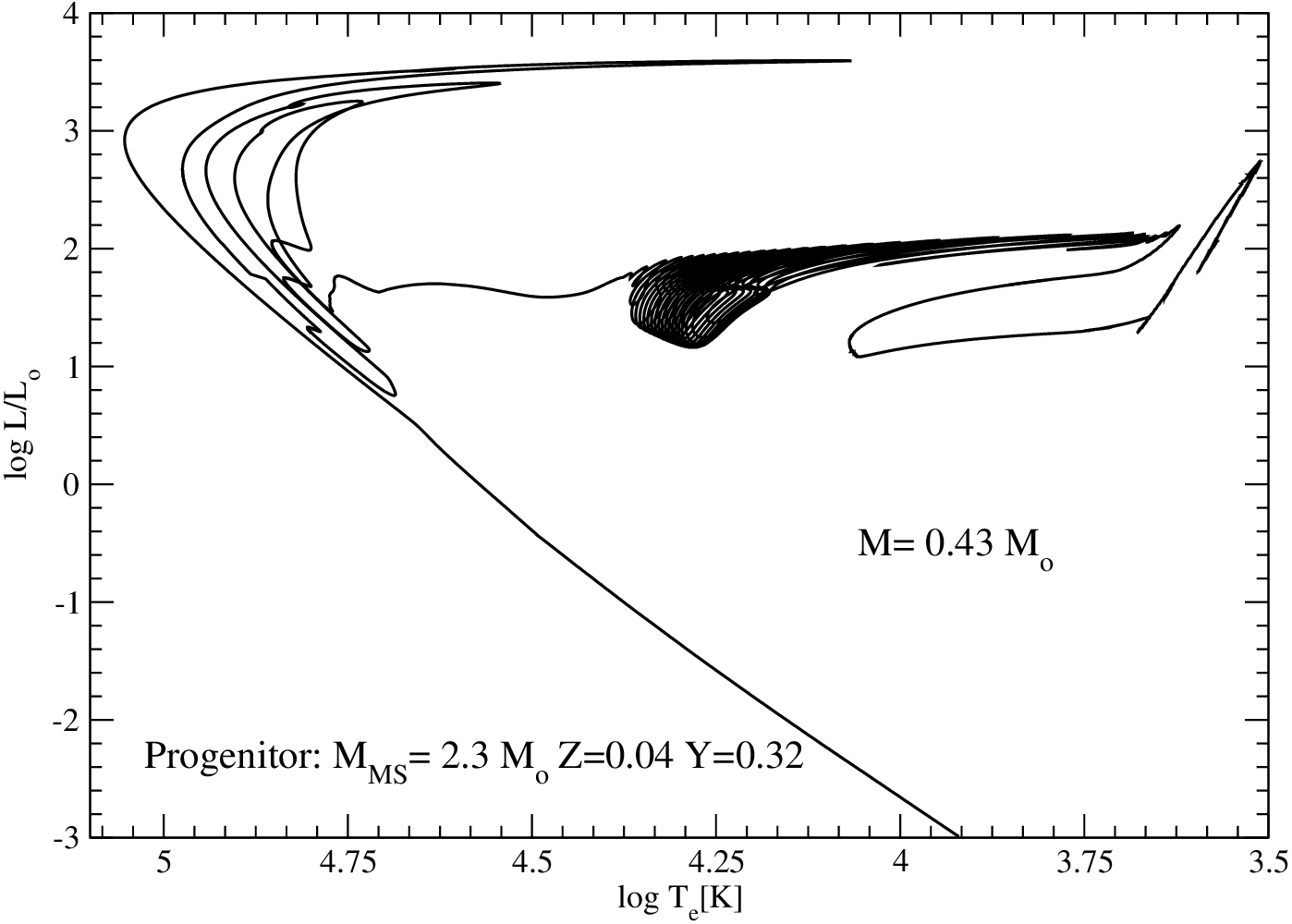}
\caption{\label{M0p43}Same as in figure \ref{M0p48} but for a remnant of mass M=0.43 M$_{\odot}$.}
\end{minipage}\hspace{2pc}%
\begin{minipage}{16pc}
\includegraphics[width=16pc]{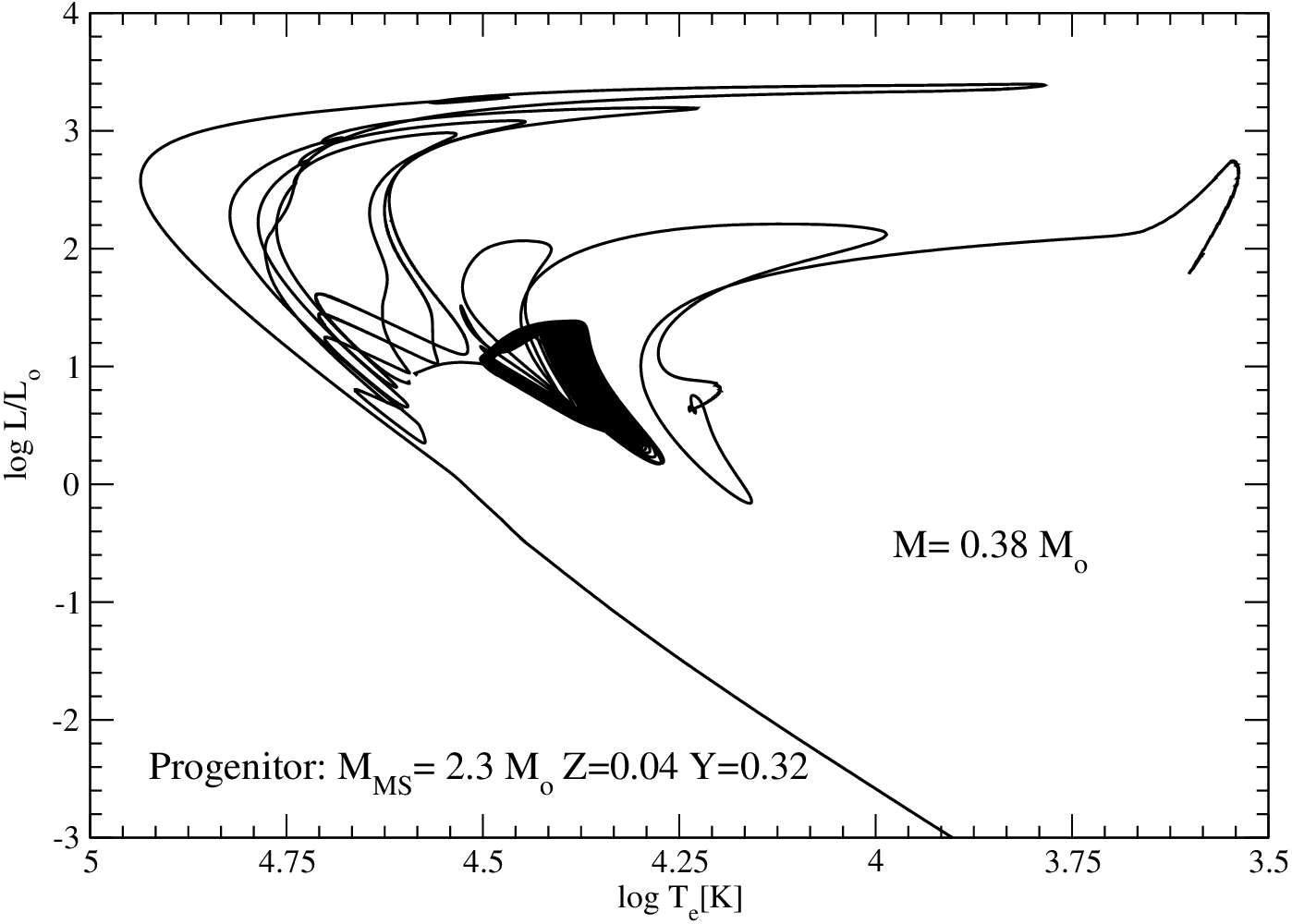}
\caption{\label{M0p38}Same as in figure \ref{M0p48} but for a remnant of mass M=0.38 M$_{\odot}$.}
\end{minipage}
\end{figure}

\begin{figure}[h]
 \begin{minipage}{16pc}
\includegraphics[width=16pc]{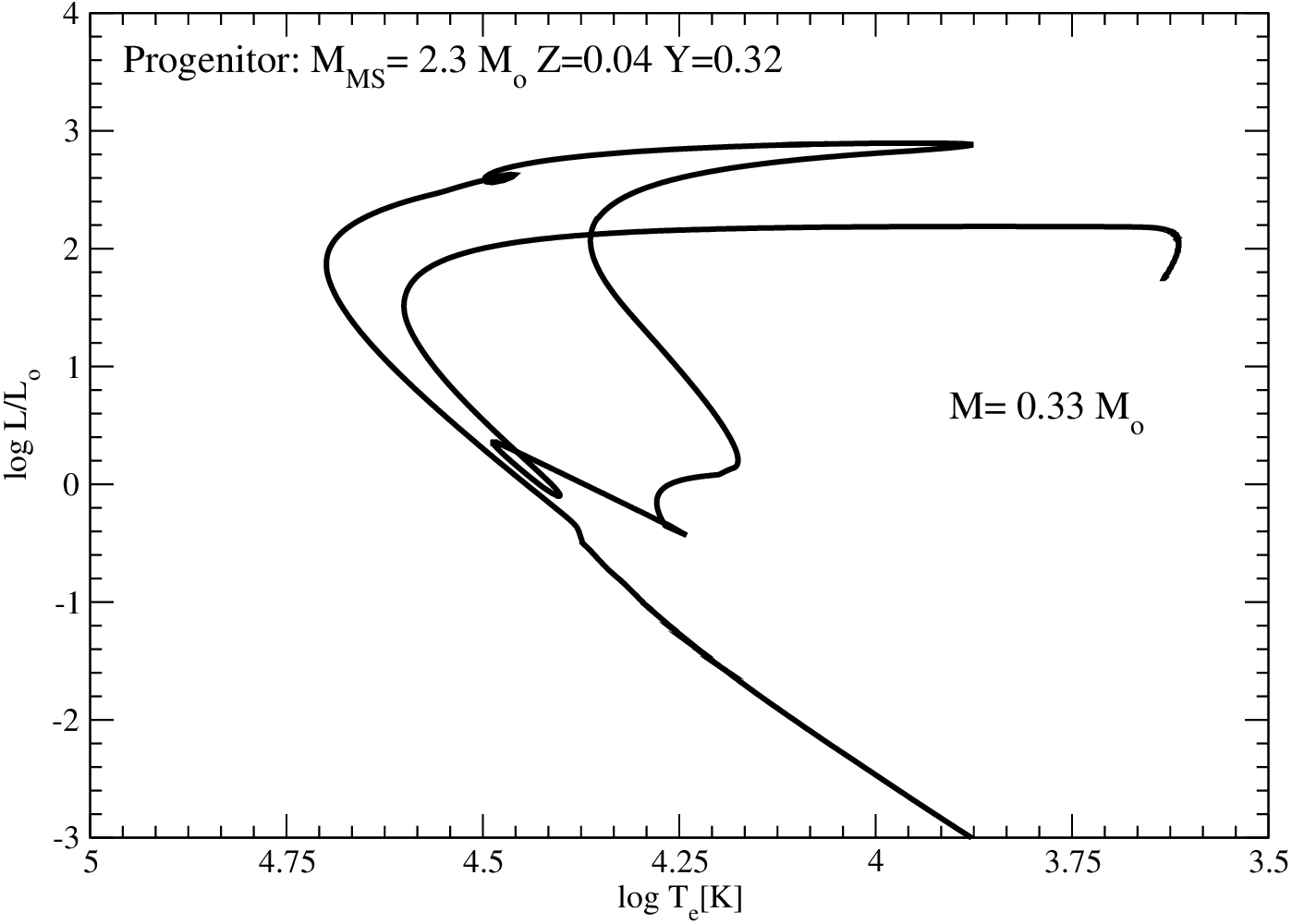}
\caption{\label{M0p33a}Same as in figure \ref{M0p48} but for a remnant of mass
 M=0.33 M$_{\odot}$. The model does not succeed to ignite 3$\alpha$ and
 eventually cools as a He-core WD.}
\end{minipage}\hspace{2pc}%
\begin{minipage}{16pc}
\includegraphics[width=15pc]{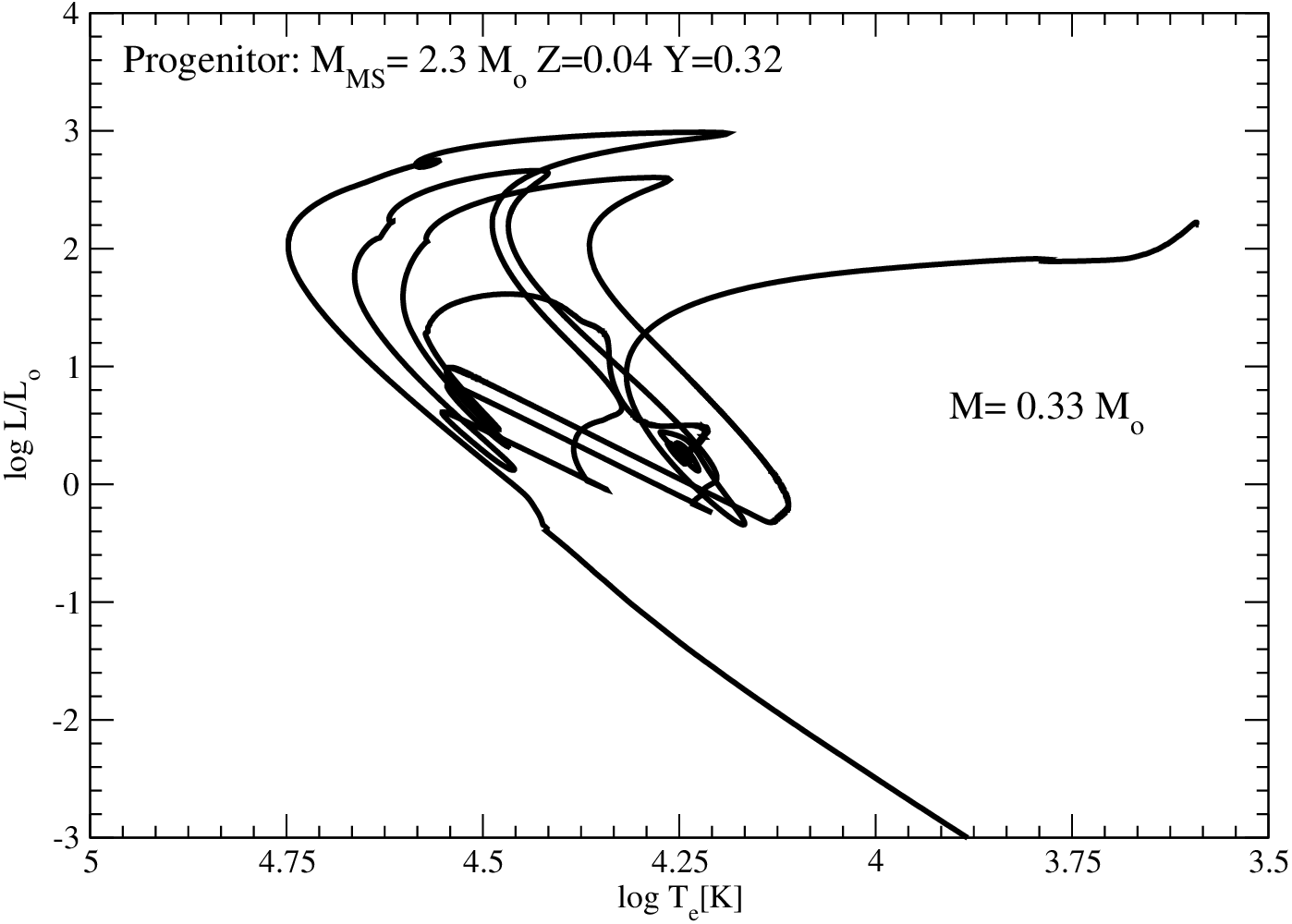}
\caption{\label{M0p33b}Same as in figure \ref{M0p48} but for a remnant of mass
 M=0.33 M$_{\odot}$. In this case the model ignites He-burning and
 eventually cools as a C/O-core WD.}
\end{minipage}
\end{figure}

The figures \ref{M0p48} - \ref{M0p33a} show the evolutionary tracks in the
HR-diagram for different values of the final mass, from 0.48 M$_{\odot}$ to
0.33 M$_{\odot}$, from the red giant to the final WD phase.
The model showed in figure \ref{M0p48}, with a total mass of 0.48 M$_{\odot}$, 
succeded to ignite He-burning and become a WD with a C/O core whose mass
fraction is about 89 \%, where we defined the core edge to be the point
 where the helium abundance became larger than 50\%.
 A very similar evolution is followed by the model
with 0.461 M$_{\odot}$, as can be seen in figure \ref{M0p461}. 
In this case, the resulting WD has a C/O core of 85\% of the total mass.
\vspace{2pc}

The evolution becomes more complex and difficult to compute for lower masses. 
The figure \ref{M0p43} shows the evolution of the model with 0.43 M$_{\odot}$. 
The star onsets the 3$\alpha$ nuclear reaction and after the central
He-burning phase it experiences a series of He-shell thermal pulses: during each
flash the star describes a loop in the HR diagram.
After the end of the thermal pulses the helium-burning shell
continues quiescently and the model moves toward higher effective tempearures. 
Then the star, approaching the cooling track, experiences
a few strong hydrogen-shell flashes, more specifically CNO-flashes. During these
episodes, the hydrogen rich outermost layer is progressively eroded, until the 
star can cool down as a WD with a C/O core of 79\% of the total mass. 
The evolution of the model with 0.38 M$_{\odot}$, shown in figure \ref{M0p38},
presents again both the He-thermal pulses and the CNO-flashes. In addition,
after the end of themal pulses, when the model moves toward the blue it
experiences a late thermal pulse.
The resulting WD has a C/O core of 50\% of the total mass.

Notice that the lower the mass of the remnant, the lower the mass fraction 
of the C/O core, thus the larger the mass of the helium-rich buffer. 
In standard C/O WD, the thickness of the He-rich layer is of the order of 
a few percent (1-2\%), while in the present models of very low mass C/O WDs
it can reach the 50\% of the total mass.

Notice that the helium-shell thermal pulses we 
showed above are not the same that occur in AGB stars. 
In fact, in that case the thermal instability is the consequence
of the accumulation of a critical mass of helium accreted by the
 quiescent hydrogen-burning shell. 
While in this case, they are self-excited relaxation oscillations, 
where the compression is due to the relaxation of the star after  
a previous contraction. 
At variance with thermal pulses in AGB, these pulses affect 
the entire star, with large oscillations in central temperature
 and density, as early shown in the pioneering paper by Iben et al. (1986).

The evolution of the remntant with mass M= 0.33 M$_{\odot}$ deserves
 a detailed discussion. 
In fact, adopting the same mass loss rate along the red giant phase of the
progenitor star of 2.3 M$_{\odot}$ as in the previous cases,  
 the 0.33 M$_{\odot}$ does not experience 3$\alpha$ burning and becomes a 
 WD with a helium core. Figure \ref{M0p33a} shows the related evolutionary track on
 the HR diagram. 
On the other hand, if the mass loss is turned on later on the RGB phase of the
progenitor star, when the mass of the helium core is larger ($\Delta M_{He} \approx$ 1\%), the final model
 with 0.33 M$_{\odot}$ succeeds to ignite He-burning and eventually evolves to
 the final WD stage with a C/O core of about the 53\% of the total mass.
 Figure \ref{M0p33b} shows its evolutionary track. 
This is the lowest C/O WD we managed to produce for this chemical composition,
in fact the models with mass smaller than 0.33 M$_{\odot}$ do not ignite 
the 3$\alpha$ and finish their evolution as He-core WDs.
The present evolutionary computations prove that it is possible 
to have C/O WDs with mass as low as 0.33 M$_{\odot}$, significantly lower
 than 0.5 M$_{\odot}$, the classical and commonly accepted lower limit. 
This means that in the mass range 0.33 - 0.5 M$_{\odot}$ both He and C/O core 
WDs can exist. Thus it is interesting to compare the main characteristics 
of their structure and evolution. A detailed description of the 
cooling evolution of low mass WDs has been recently published by Panei et al. (2007). 
\begin{figure}[h]
\begin{minipage}{16pc}
\includegraphics[width=16pc]{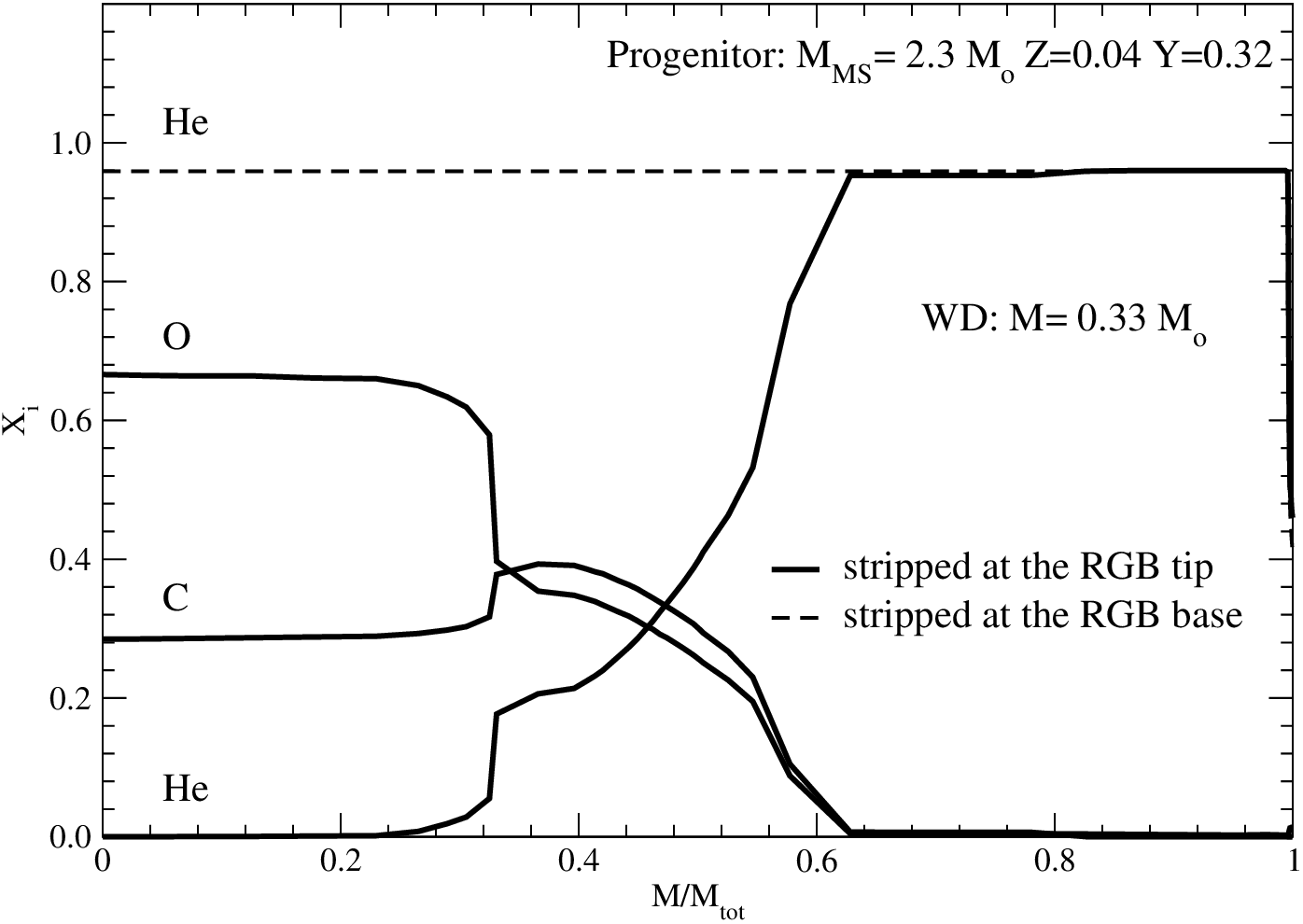}
\caption{\label{profili} Abundances in mass of helium, carbon and oxygen 
as a function of the mass coordinate for the two WDs with M= 0.33 M$_{\odot}$,
 the He-core (dashed line) and the C/O core (solid lines).}
\end{minipage}\hspace{2pc}%
\begin{minipage}{16pc}
\includegraphics[width=16pc]{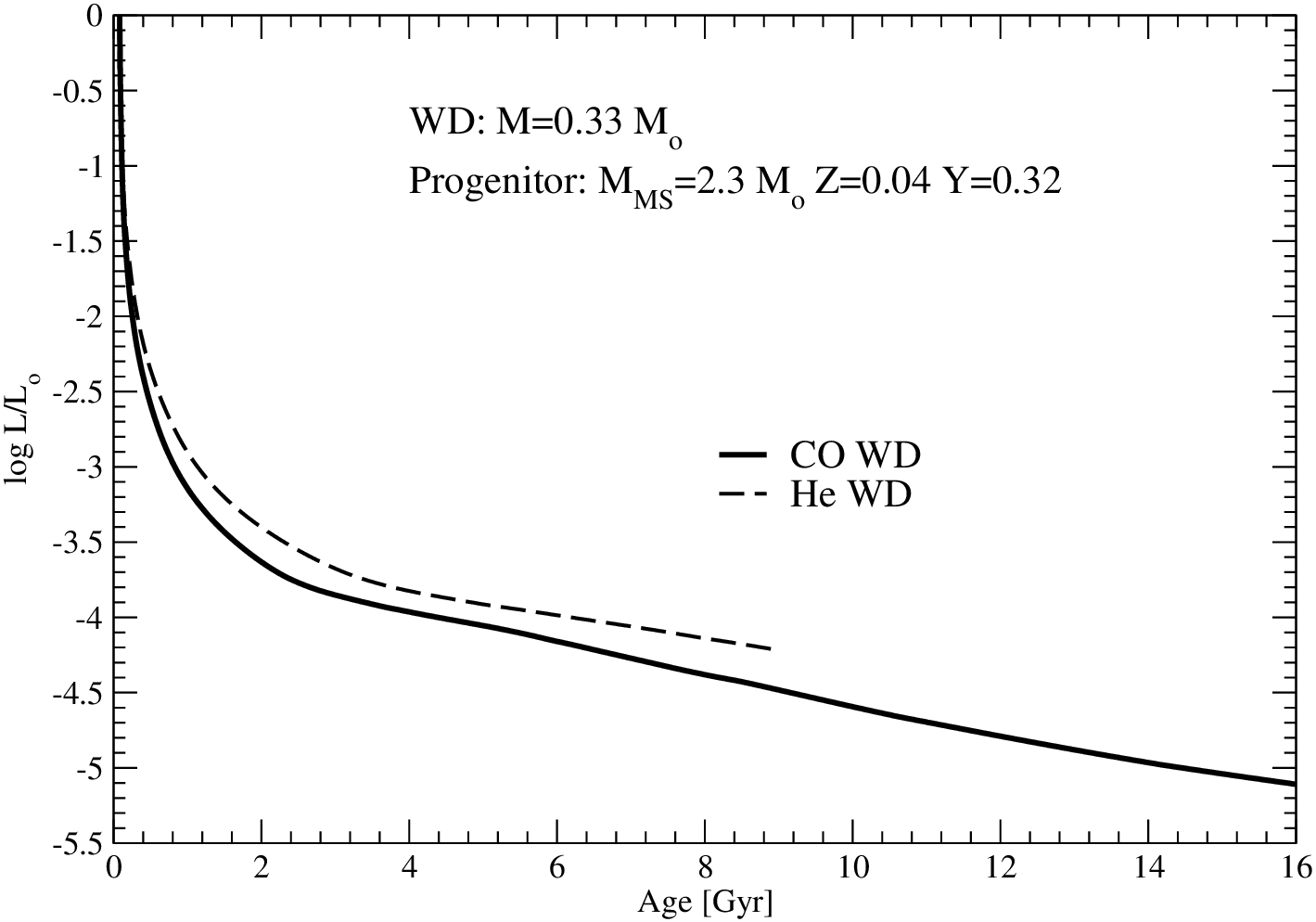}
\caption{\label{cooling} Comparison between the cooling curves,
  logL/L$_{\odot}$ vs. time, of the two WDs of 0.33 M$_{\odot}$,
that with the He-core (dashed line) and with C/O core (solid line).}
\end{minipage}\vspace{2pc}
\begin{minipage}{16pc}
\includegraphics[width=16pc]{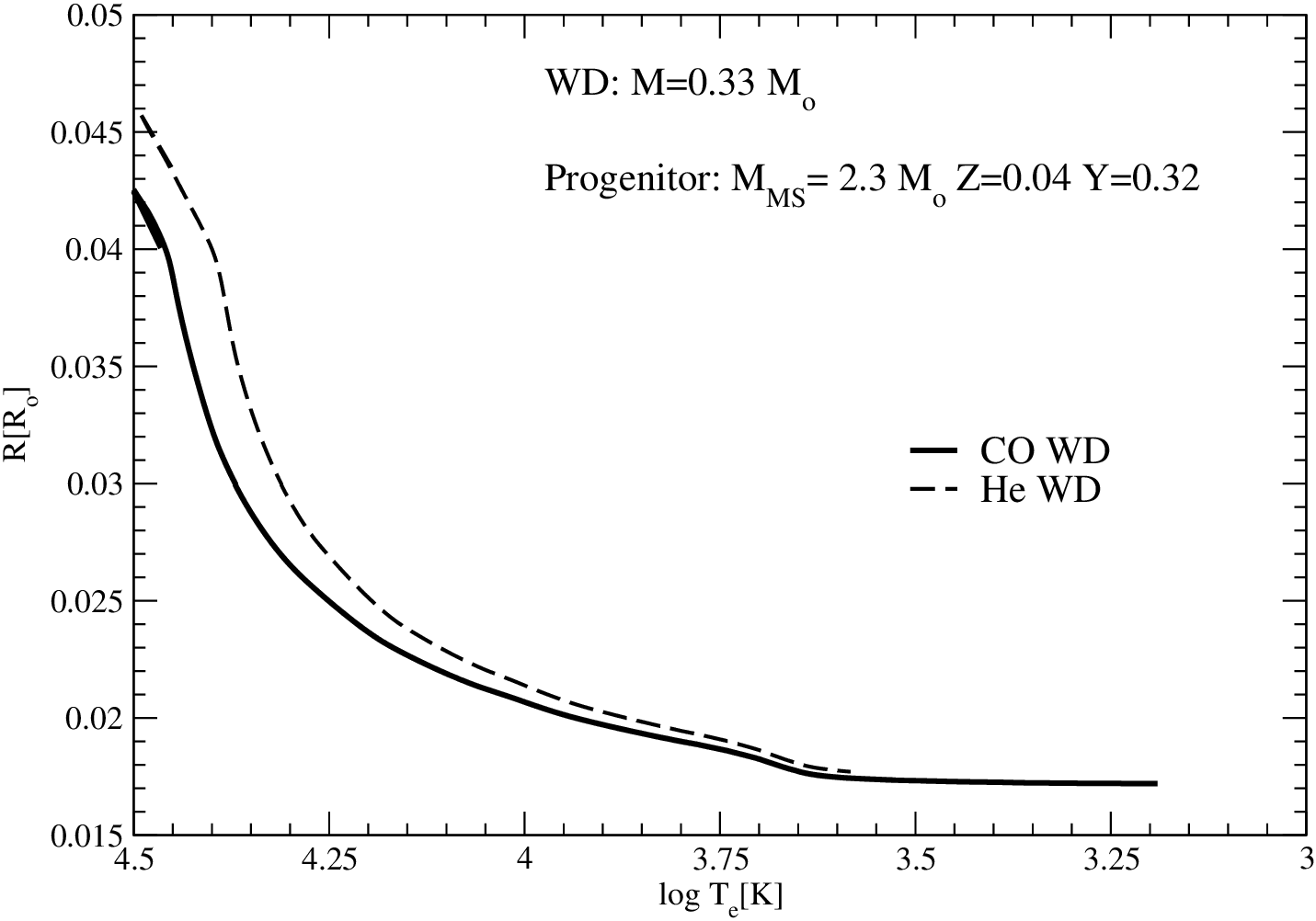}
\caption{\label{RTe}Radius vs. effective temperature 
 of the two WDs of 0.33 M$_{\odot}$, that with the He-core (dashed line)
 and with C/O core (solid line).}
\end{minipage}\hspace{2pc}%
\begin{minipage}{16pc}
\includegraphics[width=16pc]{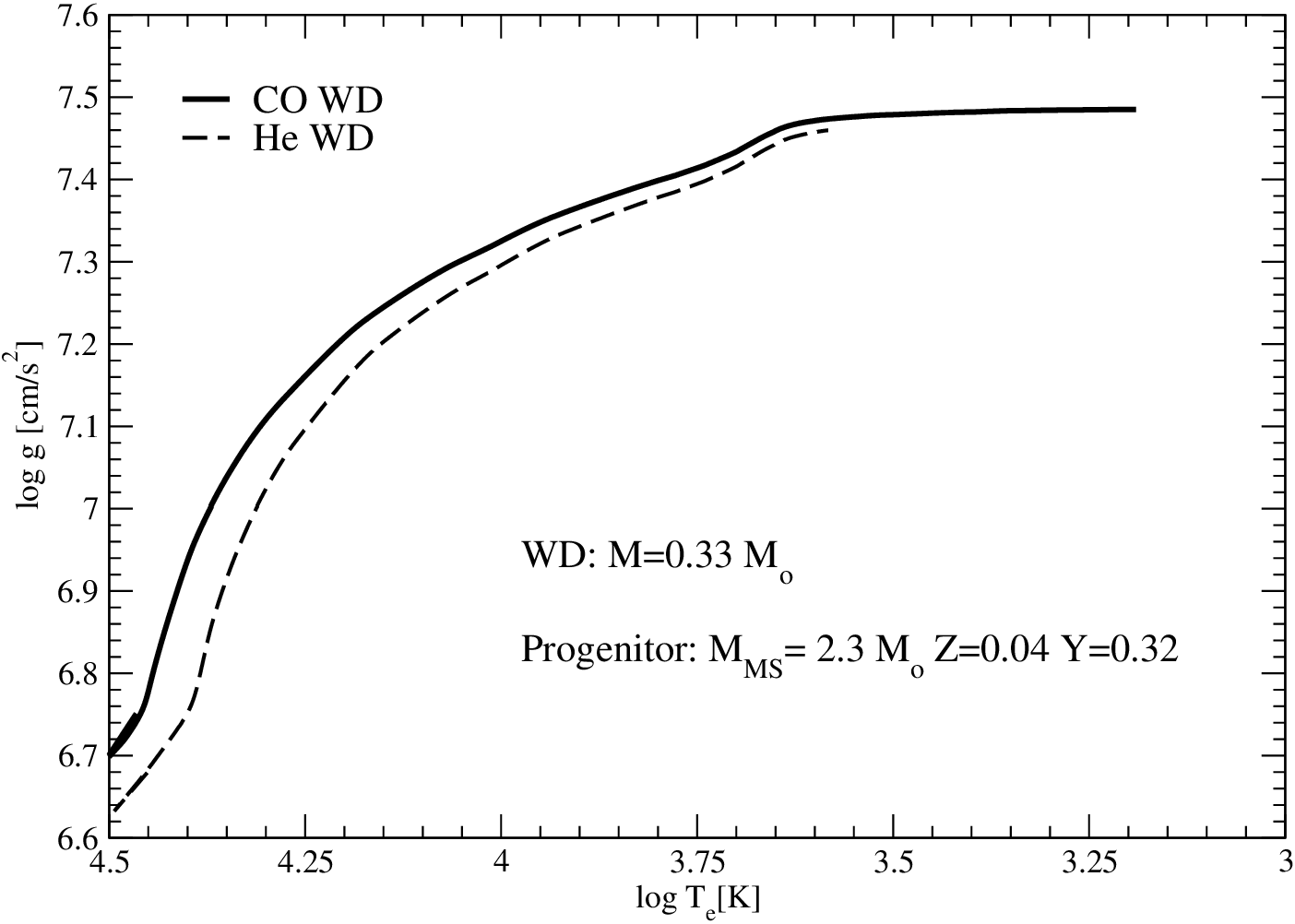}
\caption{\label{GTe}Surface gravity vs. effective temperature 
 of the two WDs of 0.33 M$_{\odot}$, that with the He-core (dashed line)
 and with C/O core (solid line).}
\end{minipage}
\end{figure}
Figure \ref{profili} shows the comparison between the core chemical 
profiles of the two remnants of 0.33 M$_{\odot}$. 
While in one case the core is constituted by nearly pure helium, in the 
other a mixture of carbon and oxygen is present. Notice that in the center
there is a significant abundance of carbon, X$_{12C}$= 0.285, as in standard 
C/O WDs, thus these very low mass WDs should not be called simply oxygen WDs.
As previously stated, a peculiar feature of these very low mass 
C/O WDs is that the helium-rich buffer is very thick, up to about 50\% of
 the total mass, while the value of typical C/O WDs is of the order
 of 1-2 \%. The thickness of the hydrogen rich outermost layer is 
almost the same in the two models, that is M= 0.0014 M$_{WD}$ for
 the WD with a helium rich core and M= 0.0015 M$_{WD}$ for that with C/O one.

Figure \ref{cooling} shows the comparison between the cooling curves
of the He-core WD (dashed line) and the C/O core (solid line). 
As expected, the remnant with the helium rich core cools slower
 than that with a C/O one, as the specific heat per gram of helium 
is larger than that of carbon and oxygen; thus the thermal content
 of the He WD is larger than that of a C/O one with the same total mass.
Such a difference in the cooling times is slightely counterbalanced 
at $logL/L_{\odot} \approx$ -4 as the C/O core crystallizes, with 
the concomitant energy release, while the He core does not, in the 
presently computed range of temperatures and densities.

Figures \ref{RTe} and \ref{GTe} show the comparison between the
 radii and surface gravities, respectively, as a function of
 the effective temperature for the two WDs of 0.33 M$_{\odot}$. 
For a given effective temperature, the He-core WD is more expanded
 than the other one, difference not negligible at the beginning of
 the cooling ($\sim$ 8.5\%).
\begin{figure}[h]
\begin{minipage}{16pc}
\includegraphics[width=16pc]{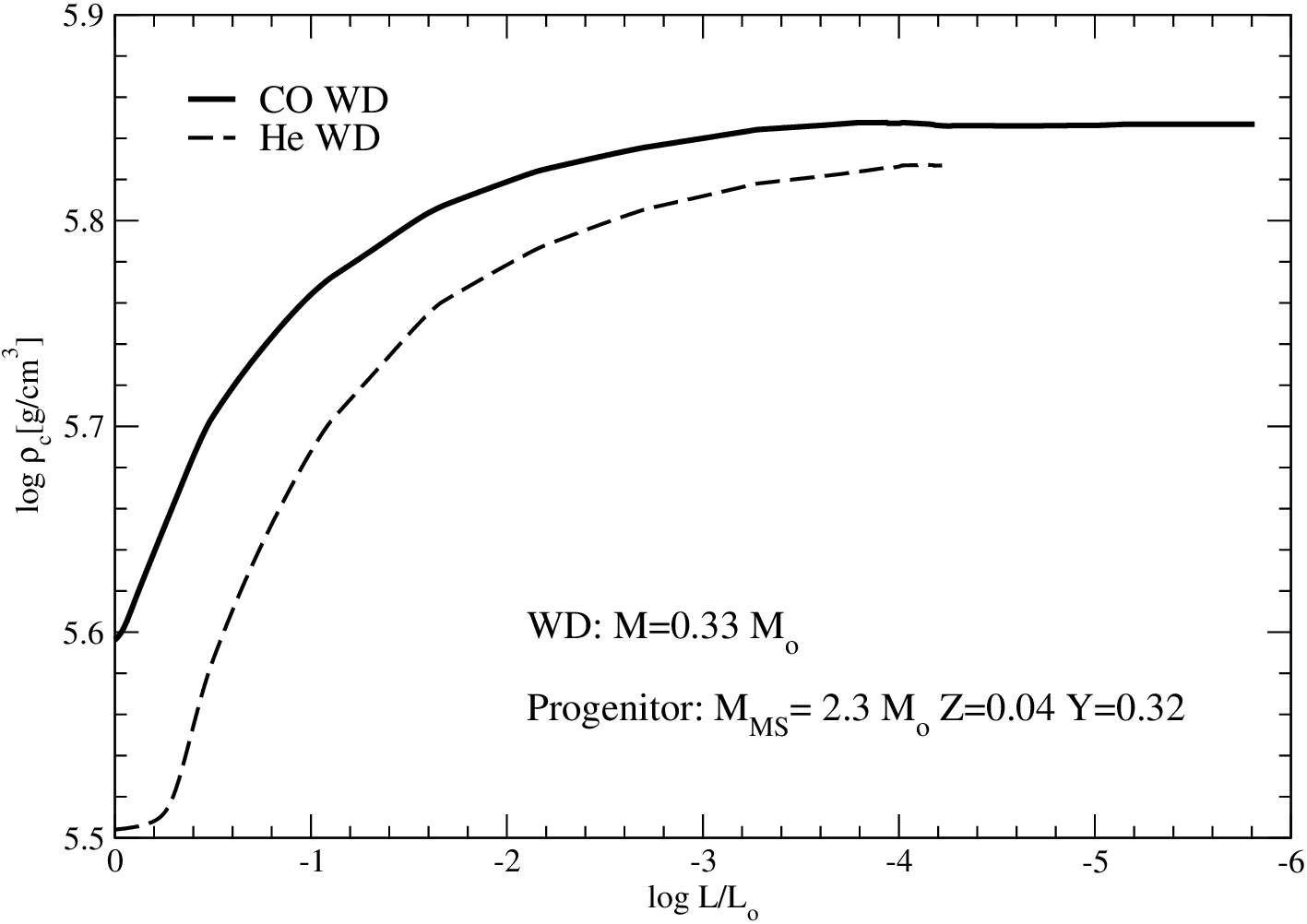}
\caption{\label{RoL}Central density $\rho_c$ vs. luminosity
 of the two WDs of 0.33 M$_{\odot}$, that with the He-core (dashed line)
 and with C/O core (solid line).}
\end{minipage}
\end{figure}
Figure \ref{RoL} shows the comparison between the
 central density $\rho_c$ as a function of the luminosity
 for the two WDs of 0.33 M$_{\odot}$, 
with the He-core (dashed line) and C/O one (solid line). 
We did not show the plot with the central temperatures T$_c$ versus
luminosity relations for the two WDs because they are nearly identical. 
\section{Conclusion}
In conclusion, as already anticipated by Iben \& Tutukov (1985), we proved, 
by means of fully and consistent evolutionary computations, that 
the minimum mass for a C/O WD is about 0.33 M$_{\odot}$, near the minimum 
possible mass of the helium-core at the onset of He-burning, which occurs at 
the RGB phase transition (M$\approx$ 2.3 M$_{\odot}$).

As a consequence, in the mass range 0.33-0.5 M$_{\odot}$ both He and C/O core
 WDs can exist. 
As expected and already shown by Panei et al. (2007), the cooling times of these two
 classes of WDs are quite different, being the He-core remnants significantly 
slower than the C/O ones. In addition, we showed also that the He WDs are 
more expanded that their C/O counterparts at a given effective temperature.
\subsection{Acknowledgments}
It's a pleasure to thank Giuseppe Bono, who kindly read the paper, for
the many useful and pleasant discussions.

\section*{References}
\begin{thereferences}
\item Castellani V, Chieffi S \& Straniero O 1992 {\it ApJSS} {\bf 78} 517
\item Catalan S, Isern J, Garcia-Berro E \& Ribas I 2008 {\it MNRAS} {\bf 387} 1693
\item Chandrasekhar S 1931 {\it ApJ} {\bf 74} 81
\item Degl'Innocenti S, Prada Moroni P G, Marconi M \& Ruoppo A 2008 {\it Ap\&SS} {\bf 316} 25
\item Dominguez I, Chieffi A, Limongi M \& Straniero O 1999 {\it ApJ} {\bf 524} 226
\item Girardi L 1999 {\it MNRAS} {\bf 308} 818
\item Hansen C J, Kawaler S D \& Trimble V 2004 {\it Stellar Interiors} (Springer)
\item Han Z, Tout C A \& Eggleton P P 2000 {\it MNRAS} {\bf 319} 215
\item Hoyle F. \& Schwarzschild M. 1955 {\it ApJS} {\bf 2} 1
\item Iben I Jr 1967 {\it ApJ} {\bf 147} 650
\item Iben I Jr, Fujimoto M Y, Sugimoto D \& Miyaji S 1986 {\it ApJ} {\bf 304} 217
\item Iben I Jr \& Tutukov A V 1985 {\it ApJSS} {\bf 58} 661
\item Meng X, Cheng X \& Han Z 2008 {\it A\&A} {\bf 487} 625
\item Panei J A, Althaus L G, Chen X \& Han Z 2007 {\it MNRAS} {\bf 382} 779
\item Prada Moroni P G \& Straniero O 2002 {\it ApJ} {\bf 581} 585
\item Prada Moroni P G \& Straniero O 2007 {\it A\&A} {\bf 466} 1043
\item Renzini A \& Buzzoni A 1983 {\it memSAIT} {\bf 54} 739
\item Stoner E C 1932 {\it MNRAS} {\bf 92} 662
\item Weidemann V 2000 {\it A\&A} {\bf 363} 647
\end{thereferences}
\end{document}